\documentclass[11pt,reqno]{article}

\usepackage{mathtools}
\usepackage{amsmath}
\usepackage{amsfonts}
\usepackage{amssymb}
\usepackage{amsxtra}
\usepackage{graphicx}
\usepackage{caption}
\usepackage{ushort}
\usepackage{color}
\usepackage{hyperref}
\usepackage[parsep]{collref}
\hypersetup{linktocpage=true}
\usepackage{enumitem}
\usepackage{titlesec}
\usepackage{cite}
\usepackage{float}
\def\nn{\nonumber} \def\bd{\begin{document}} \def\ed{\end{document}}
\def\ds{\documentstyle} \let\fr=\frac \let\bl=\bigl \let\br=\bigr
\let\Br=\Bigr \let\Bl=\Bigl
\let\bm=\bibitem
\let\na=\nabla
\let\pa=\partial \let\ov=\overline

\newcommand{\be}{\begin{equation}}
\newcommand{\ee}{\end{equation}}
\newcommand{\bea}{\begin{eqnarray}}
\newcommand{\eea}{\end{eqnarray}}
\newcommand{\ba}{\begin{array}}
\newcommand{\ea}{\end{array}}

\def\ft#1#2{{\textstyle{{\scriptstyle #1}\over {\scriptstyle #2}}}}
\def\fft#1#2{{#1 \over #2}}\def\del{\partial}
\def\vp{\varphi}
\def\sst#1{{\scriptscriptstyle #1}}
\def\st#1{{\scriptstyle #1}}
\def\oneone{\rlap 1\mkern4mu{\rm l}}
\def\td{\tilde}
\def\wtd{\widetilde}
\def\ie{{\it i.e.\ }}
\def\iec{{\it i.e.,\ }}
\def\eg{{\it e.g.\ }}
\def\egc{{\it e.g.,\ }}
\def\dalemb#1#2{{\vbox{\hrule height .#2pt
        \hbox{\vrule width.#2pt height#1pt \kern#1pt
                \vrule width.#2pt}
        \hrule height.#2pt}}}
\def\smsquare{\mathord{\dalemb{6.8}{7}\hbox{\hskip1pt}}}
\newcommand{\ho}[1]{$\, ^{#1}$}
\newcommand{\hoch}[1]{$\, ^{#1}$}
\newcommand{\ra}{\rightarrow}
\newcommand{\lra}{\longrightarrow}
\newcommand{\Lra}{\Leftrightarrow}
\newcommand{\ap}{\alpha^\prime}
\newcommand{\bp}{\tilde \beta^\prime}
\newcommand{\tr}{{\rm tr} }
\newcommand{\Tr}{{\rm Tr} }
\def\0{{\sst{(0)}}}
\def\1{{\sst{(1)}}}
\def\2{{\sst{(2)}}}
\def\3{{\sst{(3)}}}
\def\4{{\sst{(4)}}}
\def\5{{\sst{(5)}}}
\def\6{{\sst{(6)}}}
\def\7{{\sst{(7)}}}
\def\8{{\sst{(8)}}}
\def\9{{\sst{(9)}}}
\def\ten{{\sst{(10)}}}
\def\n{{\sst{(n)}}}
\def\cA{{{\cal A}}}
\def\cF{{{\cal F}}}
\def\tV{\widetilde V}
\def\tW{\widetilde W}
\def\tH{\widetilde H}
\def\tE{\widetilde E}
\def\tF{\widetilde F}
\def\tA{\widetilde A}
\def\im{{{\rm i}}}
\def\tY{{{\wtd Y}}}
\def\ep{{\epsilon}}
\def\vep{{\varepsilon}}
\def\bD{{{\bar D}}}
\def\alp{{{\a'}^3}}
\def\bD{{{\bar D}}}
\def\R{{{\mathbb R}}}
\def\C{{{\mathbb C}}}
\def\E{{{\mathbb E}}}
\def\H{{{\mathbb H}}}
\def\CP{{{\mathbb C}{\mathbb P}}}
\def\RP{{{\mathbb R}{\mathbb P}}}
\def\Z{{{\mathbb Z}}}
\def\bA{{{\mathbb A}}}
\def\bB{{{\mathbb B}}}
\def\bC{{{\mathbb C}}}
\def\bR{{{\mathbb R}}}
\def\bD{{{\mathbb D}}}
\def\bE{{{\mathbb E}}}
\def\bM{{{\mathbb M}}}
\def\bZ{{{\mathbb Z}}}
\def\Re{{{\frak{Re}}}}
\def\Im{{{\frak{Im}}}}
\def\cosec{{\,\hbox{cosec}\,}}
\def\Gm{{\Gamma_{\!\! -}}}
\def\Gp{{\Gamma_{\!\! +}}}

\def\cosech{{\hbox{cosech}}}
\def\sech{{\hbox{sech}}}
\newcommand{\cB}{\mathcal{B}}
\newcommand{\cD}{\mathcal{D}}
\newcommand{\cE}{\mathcal{E}}
\newcommand{\cH}{\mathcal{H}}
\newcommand{\cJ}{\mathcal{J}}
\newcommand{\cK}{\mathcal{K}}
\newcommand{\cL}{\mathcal{L}}
\newcommand{\cM}{\mathcal{M}}
\newcommand{\cO}{\mathcal{O}}
\newcommand{\cS}{\mathcal{S}}
\newcommand{\cU}{\mathcal{U}}
\newcommand{\cX}{\mathcal{X}}
\newcommand{\fg}{\mathfrak{g}}
\newcommand{\fh}{\mathfrak{h}}
\newcommand{\hg}{\hat{g}}
\newcommand{\hA}{{\hat{A}}}
\newcommand{\hcO}{\hat{\cO}}
\newcommand{\hcU}{\hat{\cU}}
\newcommand{\hPhi}{\hat{\Phi}}
\newcommand{\hPsi}{\hat{\Psi}}
\newcommand{\hLambda}{\hat{\Lambda}}
\newcommand{\oA}{\overline{A}}
\newcommand{\ophi}{\overline{\phi}}
\newcommand{\opsi}{\overline{\psi}}
\newcommand{\olambda}{\overline{\lambda}}
\newcommand{\tpsi}{\tilde{\psi}}

\DeclareMathOperator{\U}{U}

\newcommand{\caltech}{\it Walter Burke Institute for Theoretical Physics, California Institute of Technology, Pasadena, CA 91125}

\newcommand{\brandeis}{\it Physics Department, Brandeis University, Waltham, MA 02454}

\newcommand{\imperial}{\it The Blackett Laboratory, Imperial College London\\
Prince Consort Road, London SW7 2AZ}

\newcommand{\auth}{
C. W. Erickson\,\footnote{\,christopher.erickson16@imperial.ac.uk},
A. D. Harrold\,\footnote{\,a.harrold15@imperial.ac.uk},
Rahim Leung\,\footnote{\,rahim.leung14@imperial.ac.uk},
and K. S. Stelle\,\footnote{\,k.stelle@imperial.ac.uk}} 

\let\oldabstract\abstract
\let\oldendabstract\endabstract
\makeatletter
\renewenvironment{abstract}
{\renewenvironment{quotation}%
               {\list{}{\addtolength{\leftmargin}{1em} 
                        \listparindent 1.5em%
                        \itemindent    \listparindent%
                        \rightmargin   \leftmargin%
                        \parsep        \z@ \@plus\p@}%
                \item\relax}%
               {\endlist}%
\oldabstract}
{\oldendabstract}
\makeatother

\numberwithin{equation}{subsection}

\let\oldsection\section
\renewcommand{\section}{\renewcommand{\theequation}{\thesection.\arabic{equation}}\oldsection}
\let\oldsubsection\subsection
\renewcommand{\subsection}{\renewcommand{\theequation}{\thesubsection.\arabic{equation}}\oldsubsection}

\titlespacing*{\section}
{0pt}{5.5ex plus 1ex minus .2ex}{4.3ex plus .2ex}
\titlespacing*{\subsection}
{0pt}{5.5ex plus 1ex minus .2ex}{4.3ex plus .2ex}

\begin{document}
\setcounter{page}{0}
\thispagestyle{empty}
\begin{flushright}
\hfill{
Imperial/TP/2020/KSS/02}\\
\end{flushright} 
\vspace{20pt}

\begin{center}  

{\Large {\bf Covert Symmetry Breaking}}   

\vspace{20pt}

\auth

\vspace{7pt}
\imperial

\end{center}

\vspace{1.5cm}

\begin{abstract}

Reduction from a higher-dimensional to a lower-dimensional field theory can display special features when the zero-level ground state has nontrivial dependence on the reduction coordinates. In particular, a delayed `covert' form of spontaneous symmetry breaking can occur, revealing itself only at fourth order in the lower-dimensional effective field theory action. This phenomenon is explored in a simple model of $(d+1)$-dimensional scalar QED with one dimension restricted to an interval with Dirichlet/Robin boundary conditions on opposing ends. This produces an effective $d$-dimensional theory with Maxwellian dynamics at the free theory level, but with unusual symmetry breaking appearing in the quartic vector-scalar interaction terms. This simple model is chosen to illuminate the mechanism of effects which are also noted in gravitational braneworld scenarios.

\end{abstract}
\vfill\leftline{}\vfill
\pagebreak

\tableofcontents
\addtocontents{toc}{\protect\setcounter{tocdepth}{2}}
\newpage
\pagenumbering{arabic}
\setcounter{page}{1}
\setcounter{footnote}{0}

\section{Introduction}

This paper is about a surreptitious kind of local symmetry breaking in a lower dimensional effective field theory developed from an initial variational principle formulation of a gauge-invariant theory in a higher dimension. Surreptitious, because the symmetry breaking waits two orders in an expansion of the action in fields before it reveals itself. This phenomenon derives from a ground-state solution with nontrivial dependence on the spacetime coordinates transverse to the lower dimensions, unprotected by Killing symmetries. Given the hidden onset of such breaking at higher order in an expansion, we choose to call this `covert' symmetry breaking.

The analysis of theories with local gauge symmetries via the constraints required for consistent coupling to conserved currents has a long history in classical and quantum field theory. This has been a persistent topic in the study of gravitational theories when studied from the viewpoint of local gauge theories, with frequent comparison to the structure of Yang-Mills theories and gauge-theory couplings to symmetric matter systems. Viewing gravity as a self-coupled spin-two gauge theory with an expansion in powers of the square root of Newton's constant dates back at least to the classic ADM papers \cite{Arnowitt:1962hi}, Feynman's 1962-63 lectures on gravitation \cite{Feynman:1996kb} and in particular to papers by Weinberg \cite{Weinberg:1965rz} and Deser \cite{Deser:1969wk}. This approach has also been central to the derivation of supergravity theories \cite{Freedman:1976xh, Deser:1976eh}. The general lesson that one might wish to draw from such investigations is that once a massless field of spin one or higher is coupled consistently to symmetry currents formed from other fields, or from itself, the coupling process must  thereafter continue on in lock-step fashion order-by-order in an expansion in the corresponding coupling constant. Of course, exceptions to this general pattern can certainly exist if one includes also higher-order or higher-derivative seeds of new invariants such as $\tr(F_{\rho\sigma}\nabla^\mu\nabla_\mu F^{\rho\sigma})$ in Yang-Mills theory, and so on.

A related question is the nature of the effective theory obtained in a lower dimension in a Kaluza-Klein reduction scenario, in which modes of a higher-dimensional theory are expanded into modes of a lower dimensional theory, forming mode-towers of increasing masses.
In an expansion permitting a {\em consistent truncation}, the field equations of the higher modes may be satisfied when those modes are set to zero, yielding a dimensionally reduced theory of the lowest ``zero-level'' modes alone. However, consistent-truncation reductions involve very particular structures -- \eg based upon truncation to the invariant sector under some symmetry, or more general structures such as the $S^7$ reduction of $D=11$ supergravity \cite{Freund:1980xh}. Indeed, the $S^7$ reduction of $D=11$ supergravity falls into a somewhat different category, since retention of the full zero-level $N=8$, $D=4$ gauged supergravity supermultiplet involves a reduction ansatz in which some dependence on the transverse-space coordinates is retained (angular coordinates on $S^7$ in that case). The question of consistency of that reduction has an involved history \cite{Duff:1984hn,deWit:1984nz,Duff:1985jd,deWit:1986oxb,Godazgar:2015qia}, but one important aspect of it is the existence of $SO(8)$ Killing vectors in the reduction space, coupled with unbroken gauged $N=8$ supersymmetry.

Some reductions which do not correspond to consistent truncations to lower dimensional theories are of considerable physical importance, notably reductions on compact Calabi-Yau spaces, which have no Killing symmetries. Such reductions are still in a sense ``trivial'', however, in that they involve reductions in which all dependence on the transverse-space coordinates is suppressed. Nonetheless, such Kaluza-Klein reductions are in fact technically inconsistent: the equations of motion of the non-zero-level modes can be sourced by the zero-level modes, leading to an inconsistency in setting those higher modes to zero. A proper procedure in such cases is to integrate out the higher modes instead of truncating them, and to incorporate the resulting corrections into the lower dimensional effective theory of zero-level modes. An intermediate level of consistency in some such effective theory derivations can be identified, however: one where the effects of integrating out the heavy non-zero-level modes produce only higher-derivative corrections to the effective theory of the zero-level modes. In such a case, the structure of the effective theory when approximated by retaining only a maximum of two spacetime derivatives (with higher-derivative terms suppressed by appropriate powers of the compactification-space volume) can in some cases prove to remain unchanged with respect to a standard Kaluza-Klein reduction which simply suppresses the transverse-space coordinate dependence. Examples of such  intermediate consistency to at most second-order in derivatives are the Calabi-Yau reductions of $N=2$, $D=10$ supergravity theories \cite{Duff:1989cr}.

In this paper, we consider a situation without any of the above handholds of full or second-order-in-derivatives consistency. The question we address here is motivated by an observation that one can make  in the massless effective theory of supergravity localised on a braneworld submanifold in $D=11$ supergravity \cite{Crampton:2014hia}, where the transverse space has an $H(2,2)$ hyperbolic noncompact structure \cite{Cvetic:2003xr}. This hyperbolic transverse-space structure can be used for dimensional reduction in a standard Kaluza-Klein fashion with fields independent of the transverse coordinates, but, owing to the the noncompact transverse structure, the resulting lower dimensional Newton constant vanishes. There is, however, an alternate zero-eigenvalue normalisable transverse wavefunction which can be used successfully to localise the theory in the lower dimension. Localisation to the lower dimension in that case arises because there is a {\em mass gap} between the zero-level massless fields and the massive fields which, owing to the transverse space's noncompactness, form a continuum in mass starting at the edge of the gap. The transverse-space structure of Reference \cite{Crampton:2014hia} has the additional advantage that the corresponding Sturm-Liouville problem is integrable when considered as a Schr\"odinger equation, with a potential of P\"oschl-Teller type. This opens the way to analysis of the lower-dimensional effective braneworld theory's field equations beyond linearised order, since integrals over products of the zero-mode transverse wavefunction can be done explicitly. At the quadratic order in the action, such integrals give finite normalisation factors. At the trilinear order they give a value to the effective theory's expansion constant (\ie the square root of Newton's constant) -- finite in that case owing to convergence of the relevant integrals.

The kind of puzzle which we wish to explore here arises at the very next order: cubic in the field equations, or quartic in the action. At this order, the interaction coefficient expected from the two preceding orders turns out not to have the value expected from the square of the trilinear-order expansion constant, although it is explicitly calculable and finite. This poses our key question: what happened to the gauge and diffeomorphism symmetries expected from the linearised theory's massless character and the anticipated lock-step nature of the expansion? Such problems have not heretofore been widely studied, perhaps owing to the general technical inconsistency of the reduction problem.\footnote{That the key problem starts at fourth order in expansion of the action and is unlikely to be resolved by field redefinitions has recently been highlighted in \cite{Deser:2019yig}. The integrals of general products of the hyperbolic transverse-space wavefunction were given in Reference \cite{Crampton:2014hia}, and the unanticipated values of the resulting effective-theory expansion coefficients starting at fourth order were commented upon in \cite{Stelle:2020mmg}.}

In order to confront this phenomenon in a simpler case than the hyperbolic transverse-space braneworld supergravity setting, we work here with a simpler setup: just Maxwell theory coupled to a complex scalar field and a one-dimensional transverse space which is a $z\in I=[0,1]$ line element. In order to provoke a covert symmetry-breaking structure in the effective theory one dimension lower, we impose, however, a non-standard set of boundary conditions on the fields. For the Maxwell vector field, we pick standard Dirichlet boundary conditions at the $z=0$ end of the interval $I$, but Robin boundary conditions $(\partial_z-1) A_\mu=0$ at the $z=1$ end. This causes the zero-mode transverse wavefunction to have non-trivial dependence on the transverse coordinate $z$, similarly to the dependence of the braneworld system of Reference \cite{Crampton:2014hia} on a transverse radial coordinate.

The paper is organised as follows. We work in a general higher spacetime of dimension $d+1$. In Section \ref{sec:2} we accordingly first consider pure Maxwell theory in $(d+1)$-dimensional spacetime, but with one `transverse' dimension restricted to an interval with mixed Dirichlet/Robin boundary conditions at the two ends. When expanded in terms of $d$-dimensional fields, these boundary conditions give rise to a zero-level effective theory with a transverse wavefunction linear in the $d+\hbox{first}$ dimension. In this free theory with linear field equations, however, the dynamics of the zero-level theory remains identical to that of Maxwell theory, just with a preselection of Lorenz gauge. In Section \ref{sec:3}, the discussion is then extended to an interacting $(d+1)$-dimensional model of scalar QED with the same interval and boundary conditions. The model allows for explicit evaluation of all the relevant integrals over the transverse dimension in evaluating the zero-level $d$-dimensional effective theory. It is here that we encounter the phenomenon of covert symmetry breaking. At bilinear and trilinear orders in the action, nothing untoward happens -- the trilinear level determines the effective coupling constant $e_\text{eff}$ for vector-scalar interactions. The symmetry breaking occurs at the fourth order, however: the anticipated $e_\text{eff}^2$ coefficient for vector-scalar interactions does not occur with the right coefficient. The explanation of this phenomenon lies in the surreptitious behaviour of a nonlinearly-transforming Stueckelberg field which makes its first impact only at this level. The paper ends with a Conclusion and Outlook section in which extensions of the study of this phenomenon are considered. In the Appendices, we present some details of the calculations.

\section{Maxwell on an Interval} \label{sec:2}
\label{MaxwellSection}
\noindent In this section, we shall study dimensional reduction of Maxwell theory on an interval, where the worldvolume components of the gauge field have a non-constant zero mode. Such a system arises from choosing non-standard boundary conditions. For these conditions to be incorporated into Maxwell theory consistently, the usual action needs to be augmented by a boundary term to render the variational problem well-posed. Interestingly, the variational problem only requires boundary information on the worldvolume components of the gauge field. This leads to a bifurcation of the behaviour of the worldvolume and transverse components of the gauge field on the boundary. \\
\indent To obtain a lower-dimensional theory on the Minkowski worldvolume, we substitute the generalised Fourier expansions for the components of the gauge field into both the higher-dimensional equations of motion and the higher-dimensional action. For standard $S^1$ reductions, it is known that both procedures yield the same theory. In our case, we find the same happens, but the commutativity of these procedures depends, crucially, on the addition of the boundary term in the higher-dimensional action. In other words, given that the higher-dimensional action principle is well-posed, we obtain the following commutativity diagram for higher and lower dimensional presentations: \\
\begin{figure}[H]
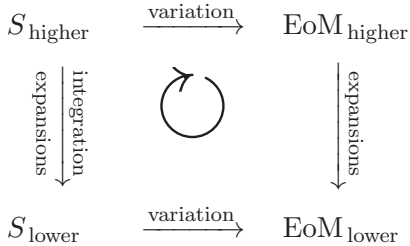

\begin{equation*}
    \begin{aligned}
        & S_{\null\,\text{higher}}
        & \xrightarrow{\text{variation}}\hspace{5mm}
        & \text{EoM}_{\text{\,higher}}\\
        & \hspace{3mm}\rotatebox[origin=c]{270}{$\xrightarrow[\text{expansions}]{\text{integration}}$}
        & \hspace{5mm}\resizebox{10mm}{7.5mm}{$\circlearrowright$}\hspace{7mm}
        & \hspace{3mm}\rotatebox[origin=c]{270}{$\xrightarrow[{\phantom{integration}}]{\text{expansions}}$}\\
        & S_{\null\,\text{lower}}
        & \xrightarrow{\text{variation}}\hspace{5mm}
        & \text{EoM}_{\text{\,lower}}
    \end{aligned}
\end{equation*}
\caption{Commuting square diagram for the reduction of dimensional presentations}\label{fig:reducsquare}
\end{figure}
\indent Since Maxwell theory is a free theory, the truncation of the lower-dimensional theory to the zero mode sector is consistent. Going back to the standard $S^1$ reductions, we recall that the zero mode sector of Maxwell theory describes a free, massless gauge field together with a massless scalar which is decoupled from the gauge field, whereas the higher modes describe massive gauge fields with masses arising from coupling to corresponding Stueckelberg scalars. In our case, we find that the theory describing the higher modes agrees with the usual $S^1$ results, but that the zero-level sector is markedly different. We will show that this sector describes a massless gauge field with an accompanying Stueckelberg scalar, which does not, however, give rise to a mass, as well as with another scalar that acts as a Lagrange multiplier imposing a Lorenz gauge condition. On-shell, this noninteracting lower-dimensional theory describes a massless photon, but it possesses one propagating degree of freedom fewer than the zero-level sector of a standard $S^1$-reduced Maxwell theory: neither the Stueckelberg scalar nor the second scalar contribute a physical degree of freedom.  The appearance of the Stueckelberg field in the zero-level sector is a direct consequence of the non-constant transverse space zero mode chosen for the worldvolume components of the gauge field. Its presence also indicates that the $\U(1)$ symmetry associated to the zero-level sector of the theory has become non-linearly realised. 
\subsection{Higher-Dimensional Equations and Boundary Conditions}\label{ssec:hdeqnsbc}
\noindent Consider Maxwell theory on a background $M_{d+1}=\bM^{1,d-1}\times I$, where $I=[0,1]$. The metric on $M_{d+1}$ will be taken to be
\begin{equation}
ds^2(M_{d+1}) = \eta_{\mu\nu}dx^\mu dx^\nu + dz^2 \,,
\end{equation}
\noindent where $x^\mu$ are the coordinates on $\bM^{1,d-1}$, and $z$ is the coordinate on the interval $I$. Consider the following modification of the usual Maxwell theory given by the action
\begin{align}
S[A_\mu, A_z] &= S_{\text{Max}}[A_\mu,A_z]+S_{\text{BT}}[A_\mu,A_z] \nonumber \\
&= \int d^dx\int_0^1 dz \Big(-\frac{1}{4}F_{\mu\nu}F^{\mu\nu} - \frac{1}{2}F_{\mu z}F^{\mu z}\Big) + \frac{1}{2}\int d^dx\, F_{\mu z}F^{\mu z}\Big\rvert_{z=1} \,, \label{eq:iac}
\end{align}
\noindent where $F_{\mu\nu} = \partial_\mu A_\nu - \partial_\nu A_\mu$, and $F_{\mu z}= \partial_\mu A_z - \partial_zA_\mu$. This action is invariant under the standard $\U(1)$ gauge transformations
\begin{equation}
A_\mu \mapsto A_\mu + \partial_\mu\Lambda \,,\quad A_z \mapsto A_z + \partial_z\Lambda \,, 
\end{equation}
\noindent for any $\Lambda=\Lambda(x,z)$. \\
\indent The variation of \eqref{eq:iac} after integrating by parts on the Minkowski boundary at infinity, where the fields $A_\mu$ and $A_z$ and their associated derivatives are assumed to vanish, is given by
\begin{align}
\delta S[A_\mu,A_z] =& \int d^dx\int_0^1dz\Big(\big(\partial_\mu F^{\mu\nu}+\partial_zF^{z\nu}\big)\delta A_\nu + \big(\partial_\mu F^{\mu z}\big)\delta A_z\Big) \nonumber \\
&+\int d^dx\, F^{\mu z}\delta A_\mu \Big\rvert_{z=0} + \int d^dx\, \Big(F^{\mu z}\big(\delta A_\mu-\partial_z\delta A_\mu\big)-\big(\partial_\mu F^{\mu z}\big)\delta A_z\Big)\Big\rvert_{z=1} \,.
\end{align}
\noindent From this, we see that the action is extremised given imposition of the Maxwell equations of motion
\begin{align}
A_\mu: &\quad \big(\Box_d + \partial_z^2\big)A_\mu - \partial_\mu\partial^\nu A_\nu - \partial_\mu\partial_z A_z = 0 \,, \label{eq:wveqn} \\
A_z: &\quad \Box_dA_z - \partial_z\partial^\mu A_\mu = 0 \,, \label{eq:zeqn}
\end{align}
\noindent subject to the Dirichlet/Robin boundary conditions on $A_\mu$
\begin{equation} 
A_\mu(x,0) = 0 \,,\quad (\partial_z - 1)A_\mu(x,1) = 0 \,, \label{eq:drbcs}
\end{equation}
\noindent where $\Box_d=\partial_\mu\partial^\mu$. It is precisely due to the boundary term in \eqref{eq:iac} that the Robin condition for the field $A_{\mu}$ can be incorporated into a well-posed variational problem. Gauge invariance of this system requires the boundary conditions on $A_\mu$ to be gauge invariant. This requirement leads to the following restrictions on the form of valid gauge parameters:
\begin{equation}
\Lambda(x,0) = c_1 \,,\quad (\partial_z - 1)\Lambda(x,1)=c_2 \,, \label{eq:residualu1}
\end{equation}
\noindent where $c_1$ and $c_2$ are constants. Our main interest will lie in the case where $c_1=c_2=0$. \\
\indent Considering only field configurations $A_\mu$ that obey the Dirichlet/Robin boundary conditions \eqref{eq:drbcs}, the action \eqref{eq:iac} is also invariant under the following transformation
\begin{equation}
A_\mu \mapsto A_\mu + \partial_\mu \Gamma \,, \quad A_z \mapsto A_z \,, \label{eq:harmonic}
\end{equation}
\noindent where $\Box_d \Gamma = 0$ and $\partial^2_z \Gamma = 0$. This is separate from the $\U(1)$ transformations, and will be called the harmonic symmetry. The boundary conditions on $A_\mu$ are only invariant under this transformation if
\begin{equation}
\Gamma(x,0) = c_3 \,,\quad (\partial_z - 1)\Gamma(x,1)=c_4 \,,
\end{equation}
\noindent Again, we will mostly be interested in the case $c_3=c_4=0$.\\
\indent Given that $A_\mu$ satisfies the Dirichlet/Robin boundary conditions, it can be expressed as a linear combination of a complete set of functions satisfying the same boundary conditions. Such a set of functions can be obtained by solving a Sturm-Liouville (SL) eigenvalue problem. From \eqref{eq:wveqn}, the natural choice for the self-adjoint SL operator is $\partial^2_z$, and the corresponding SL eigenvalue problem is
\begin{equation}
\xi_i''(z) = -\omega_i^2\xi_i(z) \,,\quad \xi_i(0) = 0 \,,\quad \xi_i'(1) - \xi_i(1) = 0 \,,
\end{equation}
\noindent where the primes indicate $z$ derivatives. The solutions to this are
\begin{equation}
\xi_0(z) =\sqrt{3}z \,, \quad \xi_i(z)= n_i\sin(\omega_i z) \,,\quad i\in\{1,2,\dots\}
\end{equation}
\noindent where $\tan\omega_i=\omega_i$ for $\omega_i>0$, and $n_i = \sqrt{2}\csc\omega_i$ are normalisation factors. These eigenfunctions are orthonormal with respect to the $L^2(I)$ inner product.

\indent With these eigenfunctions, we can write
\begin{equation}
A_\mu(x,z) = \sum_{i=0}^\infty a^{(i)}_\mu(x)\xi_i(z) \,. \label{eq:amuexp}
\end{equation}
\indent Unlike $A_\mu$, the behaviour of $A_z$ on the boundaries must be learned from the equations of motion, as the only term containing $\delta A_z$ on the boundary in the variation of the action vanishes when the equations of motion are satisfied. By substituting \eqref{eq:amuexp} into \eqref{eq:wveqn}, we have 
\begin{equation}
\sum_{i=0}^\infty\Big(\big(\Box_d - \omega^2_i\big)a_\mu^{(i)}-\partial_\mu\partial^\nu a^{(i)}_{\nu}\Big)\xi_i(z) - \partial_\mu\partial_zA_z = 0 \,,
\end{equation}
\noindent where $\omega_0= 0$. This suggests that $\partial_zA_z$ lies within the span of $\{\xi_i(z)\}$, so
\begin{equation}
\partial_zA_z(x,z) = \sum_{i=0}^\infty b^{(i)}(x)\xi_i(z) 
\end{equation}
\noindent for some coefficient functions $b^{(i)}(x)$. Integrating this expression, and noting that for $i>0$ the antiderivative of $\xi_i(z)$ is proportional to its derivative, we have
\begin{equation}
A_z(x,z) = h(x)\zeta(z)+ \sum_{i=0}^\infty g^{(i)}(x)\xi'_i(z) \,, \label{eq:azexp}
\end{equation}
\noindent where $\zeta(z) =\sqrt{3}z^2/2$ is such that $\zeta'(z)=\xi_0(z)$, and $g^{(0)}(x)$ takes the role of an integration constant for the transverse wave equation. \\
\indent The set of functions $\{\zeta(z),\xi_i'(z)\}$ is linearly independent but not $L^2(I)$ orthonormal. The second claim is easily seen by performing the requisite integrals, and to prove the first, consider the expression
\begin{equation}
c\,\zeta(z) + \sum_{i=0}^\infty f_i\xi_i'(z) = 0 \,, \label{eq:linind}
\end{equation}
\noindent for constants $c$ and $f_i$. Taking the $\partial_z$ derivative of this, we find that
\begin{equation}
c\,\xi_0(z) - \sum_{i=1}^\infty \omega_i^2 f_i\xi_i(z) = 0 \,,
\end{equation}
\noindent which by linear independence of $\{\xi_{i\ge0}(z)\}$ and the fact that $\omega_i^2\neq0$ for $i>0$ implies that $c=0$ and $f_i=0$ for $i>0$. Substituting this back into \eqref{eq:linind} gives $f_0=0$.
\subsection{Lower-Dimensional Equations and Gauge Invariance}\label{ssec:ldeqns}
\noindent To obtain the equations of motion for the component fields $a^{(i)}_\mu(x)$, $h(x)$, and $ g^{(i)}(x)$ given the equations of motion for $A_\mu$ and $A_z$, we substitute their previously derived expansions into \eqref{eq:wveqn} and \eqref{eq:zeqn}. This gives
\begin{align}
A_\mu &:\quad \Big(\Box_da^{(0)}_\mu-\partial_\mu\partial^\nu a^{(0)}_\nu - \partial_\mu h \Big)\xi_0(z) + \sum_{i=1}^\infty\Big(\big(\Box_d - \omega^2_i\big)a_\mu^{(i)}-\partial_\mu\partial^\nu a^{(i)}_{\nu}+\omega^2_i\partial_\mu g^{(i)}\Big)\xi_i(z) = 0 \,, \label{eq:amueqn}\\
A_z &:\quad \big(\Box_d h\big)\zeta(z) + \sum_{i=0}^\infty\big(\Box_d g^{(i)}-\partial^\mu a^{(i)}_\mu\big)\xi'_i(z) = 0 \,. \label{eq:azeqn1}
\end{align}
\noindent By linear independence of $\{\xi_{i\ge0}(z)\}$, \eqref{eq:amueqn} implies the following set of lower-dimensional equations
\begin{align}
&\Box_da^{(0)}_\mu-\partial_\mu\partial^\nu a^{(0)}_\nu - \partial_\mu h = 0 \,, \label{eq:loweramuzero} \\
&\big(\Box_d - \omega^2_i\big)a_\mu^{(i)}-\partial_\mu\partial^\nu a^{(i)}_{\nu}+\omega^2_i\partial_\mu g^{(i)}= 0 \,,\quad i \in \{1,2,\dots\} \,, \label{eq:loweramuheavy}
\end{align}
\noindent from which we observe that $a^{(0)}_\mu(x)$ is massless, and $a^{(i)}_\mu(x)$ are massive with masses $\omega^2_i$ implemented via the Stueckelberg fields $g^{(i)}$ for $i>0$. The $i>0$ modes then describe the massive sectors of the theory, whereas the $i=0$ modes along with $h(x)$ describe the massless sector. \\
\indent Moving on, the linear independence of $\{\zeta(z),\xi_i'(z)\}$ in \eqref{eq:azeqn1} gives
\begin{equation}
\Box_d h = 0 \,, \label{eq:peqn}
\end{equation}
\noindent and
\begin{equation}
\Box_d g^{(i)}-\partial^\mu a^{(i)}_\mu = 0 \,,\quad i\in\{0,1,\dots\} \,. \label{eq:phieqn}
\end{equation}
\noindent The lower-dimensional equations are then given by \eqref{eq:loweramuzero}-\eqref{eq:phieqn}, and are equations governing the dynamics\footnote{Note also that \eqref{eq:peqn} and the $i>0$ equations in \eqref{eq:phieqn} can be obtained by taking the divergence of \eqref{eq:loweramuzero} and \eqref{eq:loweramuheavy} respectively.} of our theory \eqref{eq:iac} after dimensionally reducing on the interval $I$.\\
\indent So far, we have been working at the level of the equations of motion, but we can ask whether the same lower-dimensional equations can equivalently be obtained by inserting the expansions of $A_\mu$ and $A_z$ directly into the action. Being careful to include both $S_{\text{Max}}$ and $S_{\text{BT}}$, the lower-dimensional action is given by \\
\begin{align}
S[a_\mu^{(i)},h,g^{(i)}] =& \int d^dx\,\Big(-\frac{1}{4}F^{(0)}_{\mu\nu}F^{(0)\mu\nu}+\frac{3}{10}\big(\partial_\mu h\big)\partial^\mu h+\big(\partial^\mu h\big)\big(\partial_\mu g^{(0)}-a_\mu^{(0)}\big)\Big) \nonumber \\
&+\sum_{i=1}^\infty\int d^dx\,\Big(-\frac{1}{4}F^{(i)}_{\mu\nu}F^{(i)\mu\nu}-\frac{1}{2}\omega_i^2(\partial_\mu g^{(i)}-a^{(i)}_\mu\big)(\partial^\mu g^{(i)}-a^{(i)\mu}\big)\Big) \,, \label{eq:lowerac}
\end{align}
\noindent where $F^{(i)}_{\mu\nu}=\partial_\mu a^{(i)}_\nu - \partial_\nu a^{(i)}_\mu$. This yields the same equations of motion, \eqref{eq:loweramuzero}-\eqref{eq:phieqn}, as those obtained via the higher-dimensional equations of motion, and so the dimensional reduction square diagram Figure \ref{fig:reducsquare} commutes. This commutativity depends crucially on the inclusion of the boundary term in the original action. From the higher-dimensional perspective, it is this term that ensures that the variational principle is well-posed. From the lower-dimensional perspective, it is this term that ensures the decoupling of the massive sectors from the massless sector. \\
\indent At this point, it is useful to consider the gauge transformations of the lower-dimensional component fields. Recall that the $\U(1)$ gauge parameter $\Lambda$ must obey the same boundary conditions as $A_\mu$, and so it can be written as a linear combination of $\{\xi_i(z)\}$ with
\begin{equation}
\Lambda(x,z) = \sum_{i=0}^\infty \lambda^{(i)}(x)\xi_i(z) \,.
\end{equation}
\noindent The harmonic symmetry parameter $\Gamma$ also obeys the same boundary conditions as $A_\mu$ with the added requirement that $\partial^2_z\Gamma = 0$, so
\begin{equation}
\Gamma(x,z) = \gamma^{(0)}(x)\xi_0(z) \,,
\end{equation}
\noindent where $\Box_d \gamma^{(0)}=0$. The $\U(1)$ transformations of $A_\mu$ and $A_z$ in terms of the component fields are
\begin{equation}
a^{(i)}_\mu(x) \mapsto a^{(i)}_\mu(x) + \partial_\mu\lambda^{(i)}(x) \,,\quad h(x) \mapsto h(x) \,, \quad  g^{(i)}(x) \mapsto  g^{(i)}(x)+\lambda^{(i)}(x) \,,\quad i \in \{0,1,\dots\} \,. \label{eq:loweru1}
\end{equation}
\indent Similarly, only $a^{(0)}_\mu$ participates in the harmonic symmetry transformation of $A_\mu$, with
\begin{equation}
a^{(0)}_\mu(x) \mapsto a^{(0)}_\mu(x) + \partial_\mu \gamma^{(0)}(x)\,,  \label{eq:lowerharmonic}
\end{equation}
\indent From these $\U(1)$ transformations, we observe that $ g^{(i)}(x)$ is a Stueckelberg field associated to $a_\mu^{(i)}$, whereas $h(x)$ is inert. The appearance of Stueckelberg fields is not new in dimensional reductions, but what is rather non-standard here is that there is also a Stueckelberg field accompanying the massless vector $a^{(0)}_\mu$. To understand this more, we need to analyse the lower-dimensional equations of motion. Since the massive sectors decouple from the massless sector, the analysis will be done in two parts. \\\\
\noindent\textbf{1. Massive Sectors}: \\\\
\noindent The massive sectors are decoupled from each other in the noninteracting theory, and each sector is described by an action \\
\begin{equation}
S^{(i)}[a^{(i)}_\mu, g^{(i)}]= \int d^dx\,\Big(-\frac{1}{4}F^{(i)}_{\mu\nu}F^{(i)\mu\nu}-\frac{1}{2}\omega_i^2(\partial_\mu g^{(i)}-a^{(i)}_\mu\big)(\partial^\mu g^{(i)}-a^{(i)\mu}\big)\Big) \,,\quad i \in \{1,2,\dots\} \,.
\end{equation}
\noindent Here, $a^{(i)}_\mu$ is a massive spin-1 field with mass $\omega_i^2$, and $ g^{(i)}$ is its associated Stueckelberg field. The number of physical degrees of freedom is $d-1$. \\\\
\noindent\textbf{2. Massless Sector}: \\\\
\noindent The zero-level massless sector is described by the action 
\begin{equation}
S[a_\mu,h, g] = \int d^dx\,\Big(-\frac{1}{4}F_{\mu\nu}F^{\mu\nu}+\frac{3}{10}\big(\partial_\mu h\big)\partial^\mu h+\big(\partial^\mu h\big)\big(\partial_\mu g-a_\mu\big)\Big) \,, \label{eq:masslessac}
\end{equation} 
\noindent where for brevity, the superscript $(0)$ has been removed. To diagonalise the scalar kinetic terms, consider the field redefinition
\begin{equation}
\varphi_1 = k\Big( g+ \frac{3-5\sqrt{5}}{10}h\Big) \,, \quad \varphi_2 = k\Big( g + \frac{3+5\sqrt{5}}{10}h\Big) \,,
\end{equation}
\noindent where $k=5^{-\frac{1}{4}}$. From \eqref{eq:loweru1}, these transform under $\U(1)$ as
\begin{equation}
\varphi_1(x) \mapsto \varphi_1(x) + k\lambda(x) \,,\quad  \varphi_2(x) \mapsto \varphi_2(x) + k\lambda(x) \,, \label{eq:u1stueckelberg}
\end{equation}
\noindent and in terms of these variables, \eqref{eq:masslessac} reads
\begin{equation}
S[a_\mu,\varphi_1,\varphi_2] = \int d^dx\,\Big(-\frac{1}{4}F_{\mu\nu}F^{\mu\nu}-\frac{1}{2}\big(\partial_\mu\varphi_1\big)\partial^\mu\varphi_1+\frac{1}{2}\big(\partial_\mu\varphi_2\big)\partial^\mu\varphi_2 + k\big(\partial^\mu\varphi_1-\partial^\mu\varphi_2\big)a_\mu\Big) \,.\label{eq:massless4dsystem2stueckelbergs}
\end{equation}
\indent The positive sign in the kinetic term of $\varphi_2$ appears to suggests that it is a ghost. It seems odd that the lower-dimensional theory could contain a ghost, since the higher-dimensional Maxwell theory is ghost-free. However, \eqref{eq:u1stueckelberg} tells us that one of $\varphi_1$ and $\varphi_2$ is pure gauge under the $\U(1)$ symmetry, so we can always choose the gauge where $\varphi_2=0$, meaning that the theory is ghost-free. To see this more clearly, consider a further field redefinition
\begin{equation}
\Psi_1 = \varphi_1-\varphi_2 \,,\quad \Psi_2 = \varphi_1+\varphi_2 \,.
\end{equation}
\noindent These transform under $\U(1)$ as
\begin{equation}
\Psi_1(x) \mapsto  \Psi_1(x) \,,\quad \Psi_2(x) \mapsto \Psi_2(x) + 2k\lambda(x) \,.
\end{equation}
\noindent Choosing the gauge $\Psi_2=0$ and integrating by parts, the action becomes
\begin{equation}\label{eq:lorenzgaugemaxwell}
S[a_\mu, \Psi_1] = \int d^dx\,\Big(-\frac{1}{4}F_{\mu\nu}F^{\mu\nu}-k \Psi_1\big(\partial^\mu a_\mu\big)\Big) \,.
\end{equation}
\indent The scalar $\Psi_1$ is non-dynamical and acts as a Lagrange multiplier imposing the Lorenz gauge condition $\partial^\mu a_\mu = 0$. Although there is no residual $\U(1)$ gauge symmetry left after imposing the $\Psi_2=0$ gauge, there is still the harmonic symmetry \eqref{eq:lowerharmonic} which remains unbroken. It is interesting to note that the harmonic symmetry acts here exactly like the radiation-gauge residuum of Lorenz gauge in usual Maxwell theory. The Lorenz gauge condition along with the harmonic symmetry removes 2 degrees of freedom from $a_\mu$, so that the total number\footnote{A more detailed degrees of freedom count is given in Appendix A.} of physical degrees of freedom of the massless sector is $d-2$. Physically, the zero-level massless sector is identical to Lorenz-gauge Maxwell theory in $d$ dimensions.
\subsection{Orthonormality and Interactions} 
\noindent Up to this point, our work has been centred around two expansion bases: $\{\xi_i(z)\}$ and $\{\zeta(z),\xi_i'(z)\}$. The first basis is $L^2(I)$ orthonormal, as guaranteed by the Sturm-Liouville theorem, but the second is not. The lack of orthonormality in the second basis did not present a problem so far because the lower-dimensional equations were obtained from the higher-dimensional ones via linear independence alone. However, when interactions are added, the higher-dimensional equations are no longer linear. In this case, we are required to expand such terms into our chosen bases. \\
\indent In anticipation of interactions, consider using an $L^2(I)$ orthonormal basis $\{\psi_\alpha(z)\}$ instead of $\{\zeta(z),\xi_i'(z)\}$ for our noninteracting Maxwell example. For brevity, summations over the basis labels will be suppressed. The functions $\psi_\alpha(z)$ can be obtained from $\zeta(z)$ and $\xi_i'(z)$ by the Gram-Schmidt procedure, and we can write
\begin{equation}
\zeta(z) = b_\alpha\psi_\alpha(z) \,,\quad \xi_i'(z) = c_{i;\alpha}\psi_\alpha(z) \,, \label{eq:onbasis}
\end{equation}
\noindent for some constants $b_\alpha$ and $c_{i;\alpha}$. With this, \eqref{eq:azexp} becomes
\begin{equation}
A_z(x,z) = \Big(b_\alpha h(x) + c_{i;\alpha}g^{(i)}(x)\Big)\psi_\alpha(z) \coloneqq \chi_\alpha(x)\psi_\alpha(z) \,. \label{eq:aznew}
\end{equation}
\noindent This shows that from a lower-dimensional perspective, the difference between using the $\{\zeta(z),\xi_i'(z)\}$ basis and the $\{\psi_\alpha(z)\}$ basis is a set of algebraic field redefinitions $\{h(x),g^{(i)}(x)\}\leftrightarrow\{\chi_\alpha(x)\}$. It is now crucial that substituting this new expansion into the higher-dimensional equations of motion and action yields the same lower-dimensional equations, since algebraic field redefinitions do not change the physics. Since this only affects the $A_z$ sector, we only need to check the $A_z$ equation. \\
\noindent At the level of the higher-dimensional equations, substituting \eqref{eq:aznew} into \eqref{eq:zeqn} gives
\begin{equation}
\Box_d\chi_\alpha - c_{i;\alpha}\partial^\mu a_\mu^{(i)} = 0 \,, \label{eq:azneweqn1}
\end{equation}
\noindent whilst the higher-dimensional action becomes
\begin{equation}
S[a_\mu^{(i)},\chi_\alpha] = \int d^dx\,\Big(-\frac{1}{4}F^{(i)}_{\mu\nu}F^{(i)\mu\nu} - \frac{1}{2}D_{\alpha\beta}\big(\partial_\mu\chi_\alpha-c_{i;\alpha}a_\mu^{(i)}\big)\big(\partial_\mu\chi_\beta-c_{j;\beta}a^{(j)\mu}\big)\Big) \,,
\end{equation}
\noindent where $D_{\alpha\beta} = \delta_{\alpha\beta} - \psi_\alpha(1)\psi_\beta(1)$. This must be equal to \eqref{eq:lowerac}, which allows us to derive the following properties of the coefficients $b_\alpha$ and $c_{i;\alpha}$:
\begin{equation}
D_{\alpha\beta}b_\alpha b_\beta = -\frac{3}{5}\,,\quad D_{\alpha\beta}b_\alpha c_{i;\beta} = -\delta_{i0}\,,\quad D_{\alpha\beta}c_{i;\alpha}c_{j;\beta} = \delta_{ij}\omega_i^2 \,. \label{eq:contractionidentities}
\end{equation}
\noindent The equation of motion for $\chi_\alpha$ obtained from this action is
\begin{equation}
D_{\alpha\beta}\big(\Box_d\chi_\beta-c_{i;\beta}\partial^\mu a_\mu^{(i)}\big) = 0 \,, \label{eq:azneweqn2}
\end{equation}
\noindent which is equivalent to \eqref{eq:azneweqn1} if $D_{\alpha\beta}$ is invertible. To prove invertibility, note that $D_{\alpha\beta}=D(\psi_\alpha(z),\psi_\beta(z))$, where
\begin{equation}
D(f_1(z),f_2(z)) = \int_0^1dz\,f_1(z)f_2(z) - f_1(1)f_2(1) \,,
\end{equation}
\noindent and consider the set of linearly independent functions
\begin{equation}
X(z;a) = a\zeta(z) + \frac{5-3a^2}{10a}\xi_0'(z)\,,\quad Y(z;a) = a\zeta(z)-\frac{5+3a^2}{10a}\xi_0'(z)\,,\quad Z_i(z) = \frac{1}{\omega_i^2}\xi_i'(z) \,,
\end{equation}
\noindent where $a\in\mathbb{R}\setminus\{0\}$, and $i\in\{1,2,\dots\}$. In this basis, $D$ is diagonalised with $D=\text{diag}(-1,1,1,\dots)$, which means that it is invertible, and hence, \eqref{eq:azneweqn1} and \eqref{eq:azneweqn2} are equivalent.
\section{Scalar QED on an Interval} \label{sec:3}
\noindent Having seen how to dimensionally reduce Maxwell theory on an interval with a non-constant zero mode, the natural progression is to see how this can be done for an interacting gauge field. As such, we now consider the above $(d+1)$ dimensional Maxwell system coupled to a complex scalar ``matter'' field, \ie scalar QED on $\bM^{1,d-1}\times I$, with the gauge field obeying the above boundary conditions \eqref{eq:drbcs}. The boundary conditions on the complex matter scalar will be chosen to be Dirichlet/Dirichlet, as this is convenient for gauge invariance. As in the previous section, this requires augmenting the usual scalar QED action by a boundary term to ensure that the variational problem is well-posed. \\
\indent Unlike pure Maxwell theory, the interactions in scalar QED will in general couple zero modes to higher modes, so truncating to the level zero sector is now generally inconsistent. We find, in our case, that the source of this inconsistency is the non-constant zero mode. Our interest is in deriving the gauge invariant {\em effective} theory describing the zero-level sector. This is obtained by integrating out all fields whose mass is greater than or equal to the mass $\omega_1$ of the least massive gauge field. A common impression might be that the integrating-out procedure of such modes leads only to higher-derivative corrections. However, we will show that this is not the case for our system. The lowest lying mode for the complex scalar is also massive, but it is lighter than the aforementioned cutoff, so it still constitutes part of the lowest-level lower-dimensional effective theory. \\
\indent Our effective theory exhibits two novel features that are not present in standard reductions of scalar QED. In the previous section, we saw that the $\U(1)$ gauge symmetry associated to the zero-mode gauge field is non-linearly realised due to the presence of a Stueckelberg field. This is also true in the effective theory. Furthermore, we will find that the na\"{\i}vely anticipated relation between the coupling constants of the cubic and quartic interactions between the zero mode gauge field and the complex scalar is not obeyed. We will show that this seemingly covert symmetry breaking, due to the mismatch between the cubic and quartic couplings, is explained by the presence of the Stueckelberg field. Consequently, the unusual quartic coupling and the non-linear realisation of the gauge symmetry go hand-in-hand to create a nonetheless gauge invariant effective theory.
\subsection{Interacting Higher-Dimensional Equations and Boundary Conditions}\label{ssec:inthdeqnsbc}
\noindent We now turn to the effect of coupling our Maxwell system \eqref{eq:iac} to matter, which we shall take to be a complex scalar field $\Phi$ charged under the $\U(1)$ symmetry. Once again, we shall consider our theory on $\bM^{1,d-1}\times[0,1]$, and we shall take the following boundary conditions for our fields:
\begin{equation}
A_\mu(x,0) = 0 \,,\quad (\partial_z - 1)A_\mu(x,1) = 0 \,,\quad \Phi(x,0) = \Phi(x,1) = 0 \,. \label{eq:qedbcs}
\end{equation}
\noindent The action governing the dynamics of our theory is
\begin{align}
S[A_\mu, A_z,\Phi,\overline\Phi] =&\; S_{\text{SQED}}[A_\mu,A_z,\Phi,\overline\Phi] + S_{BT}[A_\mu,A_z] \nonumber \\
=&\int d^dx\int_0^1dz\,\Big(-\frac{1}{4}F_{\mu\nu}F^{\mu\nu}-\frac{1}{2}F_{\mu z}F^{\mu z} - \overline{\big(D_M\Phi\big)}D^M\Phi\Big) \nonumber \\
&+ \frac{1}{2}\int d^dx\, F_{\mu z}F^{\mu z}\Big\rvert_{z=1} \,, \label{eq:qedac}
\end{align}
\noindent where $D_M\Phi = \partial_M\Phi - ie A_M\Phi$, with $e$ the charge of the complex matter scalar. This action is invariant under the following gauge transformations:
\begin{equation}
A_\mu \mapsto A_\mu + \partial_\mu\Lambda \,,\quad A_z \mapsto A_z + \partial_z\Lambda \,,\quad \Phi\mapsto e^{ie\Lambda}\Phi \,.
\end{equation}
\noindent In order for the boundary conditions in \eqref{eq:qedbcs} to be gauge invariant, we require $\Lambda$ to obey \eqref{eq:residualu1}. \\
\indent The action is extremised given the scalar QED equations of motion
\begin{align}
A_\mu: &\quad \big(\Box_d + \partial_z^2\big)A_\mu - \partial_\mu\partial^\nu A_\nu - \partial_\mu\partial_z A_z + ie\big(\Phi\partial_\mu\overline\Phi-\overline\Phi\partial_\mu\Phi\big)-2e^2\overline\Phi\Phi A_\mu = 0 \,, \label{eq:qedwveqn} \\
A_z: &\quad \Box_dA_z - \partial_z\partial^\mu A_\mu+ie\big(\Phi\partial_z\overline\Phi-\overline\Phi\partial_z\Phi\big)- 2e^2\overline\Phi\Phi A_z= 0 \,, \label{eq:qedzeqn} \\
\Phi: &\quad\big(\Box_d + \partial_z^2\big)\Phi - ie\big(\Phi\partial^\mu A_\mu+\Phi\partial_zA_z+2A_\mu\partial^\mu\Phi + 2A_z\partial_z\Phi\big) -e^2\Phi\big(A_\mu A^\mu + A_z^2\big) = 0 \,, \label{eq:qedscalareqn}
\end{align}
\noindent subject to the boundary conditions \eqref{eq:qedbcs}. 
\subsection{Interacting Lower-Dimensional Theory} \label{ssec:intldeqnsbc}
\indent As in the previous section, the expansions for $A_\mu$ and $A_z$ are
\begin{equation}
A_\mu(x,z) = \sum_{i=0}^\infty a^{(i)}_\mu(x)\xi_i(z) \,,\quad A_z(x,z) = h(x)\zeta(z) + \sum_{i=0}^\infty g^{(i)}(x)\xi_i'(z) \,,
\end{equation}
\noindent For the complex matter scalar, we introduce another complete set of functions, $\{\theta_n(z)=\sqrt{2}\sin(m_nz)\}$ with $n\in\{1,2,\dots\}$ and $m_n = n\pi$, which satisfy Dirichlet/Dirichlet boundary conditions. Using these, the scalar field is expanded as
\begin{equation}
\Phi(x,z)= \sum_{n=1}^\infty \phi^{(n)}(x)\theta_n(z) \,.
\end{equation}
\noindent The complex scalars $\phi^{(n)}$ transform under the $\U(1)$ gauge symmetry non-diagonally with
\begin{equation}
\phi^{(n)} \mapsto \sum_{m=1}^\infty\exp\big(ie\lambda^{(i)}I_i\big)^{nm}\phi^{(m)} \,, \label{eq:scalaru1}
\end{equation}
\noindent where the matrix $I_i$ is defined as
\begin{equation}
(I_i)^{nm} = I_i^{nm} =  \int_0^1dz\,\xi_{i}(z)\theta_n(z)\theta_m(z) \,.
\end{equation}
We can now substitute the expansions of $A_\mu$, $A_z$, and $\Phi$ into the higher-dimensional equations of motion or into the higher-dimensional action to obtain a lower-dimensional theory. It is a straightforward albeit long calculation to show that both procedures give the same result, and so the Figure \ref{fig:reducsquare} dimensional reduction square once again commutes. The route involving substituting the expansions into the higher-dimensional equations is a bit subtle, and involves projecting the non-linear interaction terms into the relevant bases. For example, in \eqref{eq:qedwveqn}, we notice that the terms $\Phi\partial_\mu\overline\Phi$ and $\overline\Phi\Phi A_\mu$ obey Dirichlet/Robin conditions, and so can be written as linear combinations of the $\{\xi_i(z)\}$ basis. In particular, we have
\begin{equation}
\theta_n(z)\theta_m(z) = I_i^{nm}\xi_i(z) \,,\quad \theta_n(z)\theta_m(z)\xi_j(z)=I_{ij}^{nm}\xi_i(z) \,, \label{eq:expansionthetasxis}
\end{equation}
\noindent where summations over the index labels are suppressed, and
\begin{equation}
I_{ij}^{nm} = \int_0^1dz\,\xi_{i}(z)\xi_{j}(z)\theta_n(z)\theta_m(z)\,.
\end{equation}
\noindent We will refer the reader to Appendix B for a full treatment of the higher-dimensional equations of motion. \\
\indent To present the lower-dimensional action in a recognisable form, we define the covariant derivative operator
\begin{equation}
D^{nm}_\mu = \delta_{nm}\partial_\mu - ieI_{i}^{nm}a^{(i)}_\mu \,. \label{eq:covderiv}
\end{equation}
\noindent Using \eqref{eq:loweru1} and \eqref{eq:scalaru1} we can check that this is a proper covariant derivative with respect to the $\U(1)$ gauge symmetry, as
\begin{equation}
\big(D_\mu \phi\big)^{(n)} \mapsto \exp\big(ie\lambda^{(i)}I_i\big)^{nm}\big(D_\mu \phi\big)^{(m)} \,.\label{eq:Dtransf}
\end{equation}
\noindent Then, defining the inner product $(u,v)=\overline u^{(n)}v^{(n)}$ over the space of complex scalars, and defining the matrices $J$, $K$, and $L_i$ with components
\begin{equation}
J^{nm} = \int_0^1 dz\,\theta_n(z)\theta'_m(z) \,,\quad K^{nm} = \int_0^1dz\,\zeta(z)\theta_n(z)\theta_m(z) \,,\quad L_i^{nm} = \int_0^1dz\,\xi_i'(z)\theta_n(z)\theta_m(z) \,,\label{eq:cJKLmatrices}
\end{equation}
\noindent the lower-dimensional action becomes
\begin{align}
S=& \int d^dx\,\Big(-\frac{1}{4}F^{(i)}_{\mu\nu}F^{(i)\mu\nu}  -\frac{1}{2}\omega_i^2(\partial_\mu g^{(i)}-a^{(i)}_\mu\big)(\partial^\mu g^{(i)}-a^{(i)\mu}\big) + \frac{3}{10}\partial_\mu h\partial^\mu h \nonumber \\
&+\partial^\mu h\big(\partial_\mu g^{(0)}-a_\mu^{(0)}\big)- \big(D_\mu\phi,D^\mu\phi\big)- \big(W\phi, W\phi\big)\Big)\,, \label{eq:qedlowerac}
\end{align}
\noindent where $\omega_0^2=0$, and $W=J-iehK-ieg^{(i)}L_i$. The term $W\phi$ transforms covariantly under the $\U(1)$ transformations given in \eqref{eq:loweru1} and \eqref{eq:scalaru1} with $W\phi\mapsto UW\phi$, where $U=\exp(ie\lambda^{(i)}I_i)$. This is expected, as it is just the lower-dimensional analogue of the higher-dimensional $D_z\Phi$ term, which by definition transforms covariantly under $\U(1)$ transformations. We also note that the lowest-order term in the scalar potential $(W\phi,W\phi)$ is $(J\phi,J\phi) = m_n^2\overline\phi^{(n)}\phi^{(n)}$, which means that the lowest lying scalar $\phi^{(1)}$ is massive with mass $m_1 = \pi$. 
\subsection{An Unusual Coefficient}
\noindent The lower-dimensional action \eqref{eq:qedlowerac} containing the modes $a_\mu^{(i)}$, $h$, $g^{(i)}$, and $\phi^{(n)}$ is simply a rewriting of the higher-dimensional action \eqref{eq:qedac} in a particular choice of bases. Our goal is now to build a gauge invariant effective theory from the lower-dimensional action containing only $a^{(0)}$, $h$, $g^{(0)}$ and $\phi^{(1)}$ after integrating out the modes above level zero.\footnote{The cutoff scale is $\Lambda^2 = \omega_1^2$, noting that $\omega_1^2>m_1^2$.}  We shall show that this effective theory realises gauge invariance in a non-standard manner, notably the usual relationship between the cubic and quartic coupling constants in scalar QED is not present. In order to demonstrate this, we need to perform a set of field redefinitions on $\phi^{(n)}$ to obtain a set of fields $\varphi^{(n)}$ that transform canonically under the $\U(1)$ symmetries.\footnote{A discussion of the effective theory in the original variables is given in Appendix C.} \\
\indent From the covariant derivative operator \eqref{eq:covderiv}, we observe that the effective coupling of $\phi^{(n)}$ to each $a_\mu^{(i)}$ is $eI_i^{nn}$, with no sum over $n$. This motivates the following set of field redefinitions
\begin{equation}
\varphi^{(n)} = \exp(ieg^{(i)}I_i^{nn})\exp(-ie g^{(i)}I_i)^{nm}\phi^{(m)} \coloneqq X^{nm}\phi^{(m)} \,.\label{eq:fredefs}
\end{equation}
\noindent These transform under the $\U(1)$ symmetries as
\begin{equation}
\varphi^{(n)}\mapsto\exp(ie\lambda^{(i)}I_i^{nn})\varphi^{(n)} \,.\label{eq:U1transfsmassive}
\end{equation}
\noindent Note that $\exp\big(ieg^{(i)}I_i^{nn}\big)$ is a phase and not a matrix. The matrix $X^{nm}$ is unitary, so the mass of $\varphi^{(n)}$ is $m_n^2$. This field redefinition can be interpreted as a two-step process, each of which relies on the existence of the Stueckelberg fields, especially the zero-mode Stueckelberg, $g^{(0)}$. Since the Stueckelberg fields transform inhomogeneously by gauge parameters, we can use them to nullify or create any gauge transformation. In the case of \eqref{eq:fredefs}, we first define a set of non-transforming scalars 
\begin{equation}
\psi^{(n)}=\exp(-ie g^{(i)}I_i)^{nm}\phi^{(m)} \,. \label{eq:nontrans}
\end{equation}
\noindent Then, from this, we use the Stueckelberg fields to write down the canonically transforming scalars in \eqref{eq:fredefs}. \\
\indent The stage is now set for us to write down an effective theory of $a_\mu^{(0)}$, $h$, $g^{(0)}$, and $\varphi^{(1)}$, but before that, let's look at the portion of the theory that contains only the interactions between $a^{(0)}_\mu$ and $\varphi^{(1)}$. These terms are given by
\begin{equation}
\mathcal{L}_{\text{int}}({a^{(0)}_\mu,\varphi^{(1)}}) = -ie I_0^{11} a_\mu^{(0)}\big(\varphi^{(1)}\partial^\mu\overline\varphi^{(1)}-\overline\varphi^{(1)}\partial^\mu\varphi^{(1)}\big) - e^2I_{00}^{11}a^{(0)}_\mu a^{(0)\mu}|\varphi^{(1)}|^2 \,. \label{eq:qedintzero}
\end{equation}
\noindent As $\varphi^{(1)}$ transforms canonically under the $\U(1)$ symmetry associated with $a^{(0)}_\mu$, we might expect this to look like a standard scalar QED coupling. However, in scalar QED, the quartic coupling constant is equal to the square of the cubic coupling constant. This is not the case here, since $I_{00}^{11} \neq (I_{0}^{11})^2$. Since the full theory, given in \eqref{eq:qedlowerac}, is gauge invariant, the remedy to this unusual coefficient problem clearly lies in the modes that we have neglected. As such, we might assume that integrating out the massive vectors and heavier scalars will modify the coupling constants in \eqref{eq:qedintzero} such that the usual scalar QED structure reappears. However, this is not what happens, as we will see in the next subsection. 
\subsection{Integrating Out}\label{sc:ifo}
\noindent To integrate out the heavy modes in this theory, we will work with the assumption that the action of a massive propagator $(\Box_d - M^2)^{-1}$ acting on a current $J$ can be approximated to be
\begin{equation}
\big(\Box_d-M^2\big)^{-1}J = -M^{-2}J + \mathcal{O}(M^{-4}\Box_d) \,,
\end{equation}
\noindent Since our immediate goal is to investigate whether integrating out the massive vectors and matter scalars modifies the coefficients in \eqref{eq:qedintzero}, it is sufficient to consider only those terms in their equations of motion containing themselves, the fields $a^{(0)}_\mu$, and $\varphi^{(1)}$, a maximum of one derivative, and contributing to a cubic and a quartic interaction. Taking this into account, the relevant parts of the theory are
\begin{align}
\mathcal{L}_{\text{rel}}(a_\mu^{(i)},\varphi^{(n)}) =& -\frac{1}{2}\omega_i^2a_\mu^{(i)}a^{(i)\mu} - m_n^2|\varphi^{(n)}|^2- ieI_i^{nm}a^{(i)}_\mu\big(\overline\varphi^{(n)}\partial^\mu\varphi^{(m)}-\varphi^{(n)}\partial^\mu\overline\varphi^{(m)}\big) \nonumber \\
&- e^2I_{ij}
^{nm}a^{(i)}_\mu a^{(j)\mu}\overline\varphi^{(n)}\varphi^{(m)} \,. \label{eq:qedrel}
\end{align}
\indent From this, we find that the heavy fields are given by
\begin{align}
a^{(\underline{{i}})}_\mu &= \frac{ie}{\omega_{\underline{{i}}}^2}I_{\underline{{i}}}^{11}\big(\varphi^{(1)}\partial_\mu\overline\varphi^{(1)}-\overline\varphi^{(1)}\partial_\mu\varphi^{(1)}\big)-\frac{2e^2}{\omega_{\underline{{i}}}^2}I_{\underline{{i}}0}^{11}a_\mu^{(0)}|\varphi^{(1)}|^2 + \cdots \,, \label{eq:intoutvec} \\
\varphi^{(\underline{{n}})} &= -\frac{ie}{m_{\underline{{n}}}^2}I_0^{\underline{{n}}1}\big(\varphi^{(1)}\partial^\mu a_\mu^{(0)} + 2a_\mu^{(0)}\partial^\mu\varphi^{(1)}\big) - \frac{e^2}{m_{\underline{{n}}}^2}I_{00}^{\underline{{n}}1}a_\mu^{(0)}a^{(0)\mu}\varphi^{(1)} + \cdots \,, \label{eq:intoutscalar}
\end{align}
\noindent where $\underline{{n}}\in\{2,3,\dots\}$ and $\underline{{i}}\in\{1,2,\dots\}$, and the ellipses denote terms containing more than three fields and/or more than one derivative. Substituting \eqref{eq:intoutvec} and \eqref{eq:intoutscalar} back into \eqref{eq:qedrel}, which is allowable as the equations are algebraic, we find that the corrections are not of the same structure as in \eqref{eq:qedintzero}. This means that there is {\em no} correction to the cubic and quartic coupling constants arising from integrating out the massive fields. \\
\indent In effect, by expanding scalar QED in modes of a lower-dimensional theory, we have obtained an effective theory of a complex matter scalar coupled to a gauge field where the presence of Stueckelberg fields at all levels, including level zero, plays a crucial role in establishing gauge invariance. It is also interesting to note that, contrary to a variety of examples in the literature, integrating out the massive fields here does not solely produce higher-derivative corrections, but contributes as well to achieving gauge invariance in the lower-dimensional effective theory. For instance, the mass terms $m_{\underline{{n}}}^2\overline\varphi^{(\underline{{n}})}\varphi^{(\underline{{n}})}$ produces a sixth-order, zero-derivative correction of the form $e^4 (a^{(0)}_\mu a^{(0)\mu})^2|\varphi^{(1)}|^2/6$. 
\subsection{The Fourth-Order, Two-Derivative Effective Theory}\label{sc:tfet}
\noindent We now wish to make a full presentation of the lower dimensional effective theory after putting the heavy modes on-shell. The easiest method for this calculation is to perform the integrating out procedure in the non-transforming variables given in \eqref{eq:nontrans}, then transform back into the canonically transforming variables. In the non-transforming variables, the lower dimensional Lagrangian density takes the form:
\begin{align}
\mathcal{L} =& -\frac{1}{4}F^{(i)}_{\mu\nu}F^{(i)\mu\nu}  -\frac{1}{2}\omega_i^2(\partial_\mu g^{(i)}-a^{(i)}_\mu\big)(\partial^\mu g^{(i)}-a^{(i)\mu}\big) + \frac{3}{10}\partial_\mu h\partial^\mu h  \nonumber \\
&+\partial^\mu h\big(\partial_\mu g^{(0)}-a_\mu^{(0)}\big)- \big(\mathcal D_\mu\psi,\mathcal D^\mu \psi\big)- (\mathcal{W}\psi,\mathcal{W}\psi)\,,
\end{align}
\noindent where $\mathcal{D}_\mu^{nm} = \delta_{nm}\partial_\mu - ie\big(a_\mu^{(i)}-\partial_\mu g^{(i)}\big)I_i^{nm}$, and $\mathcal{W}=W+ieg^{(i)}L_i=J-iehK$, which is a gauge invariant quantity. \\
\indent Putting $a^{(\underline i)}$ and $\psi^{(\underline n)}$ on-shell while gauge fixing the higher-mode Stueckelberg fields $g^{(\underline i)}$ to zero, we find that the effective Lagrangian density to fourth-order in interactions and second-order in derivatives is
\begin{align}
\mathcal{L}_{\text{eff}}=&-\frac{1}{4} F_{\mu\nu}F^{\mu\nu} +\frac{3}{10}\partial_\mu h\partial^\mu h +\partial^\mu h\big(\partial_\mu g-a_\mu\big)- \partial_\mu \overline \psi \partial^\mu \psi - \pi^2 \overline \psi \psi \nonumber \\
&- e I_0^{11} \left(a_\mu - \partial_\mu g\right)\left( \overline \psi \partial^\mu \psi - \psi \partial^\mu \overline \psi\right) -e^2 I_{00}^{11}\left(a_\mu - \partial_\mu g\right)\left(a^\mu - \partial^\mu g\right)\overline \psi \psi \nonumber \\
&- e^2\left(P^{11}-\sum_{\underline n = 2}^\infty \left( \frac{T^{\underline n 1}-T^{1 \underline n}}{\pi^2 {\underline n}^2}\right)^2\right) h^2 \overline \psi \psi \;,
\end{align}
\noindent where we removed the superscripts $(0)$ and $(1)$. The overlap integrals $P^{nm}$ and $T^{nm}$ are defined in Appendix B. The coefficient of the $h^2\overline\psi\psi$ quartic interaction can be calculated exactly:
\begin{equation}\begin{aligned}
    X &= P^{11}-\sum_{\underline n = 2}^\infty \left( \frac{T^{\underline n 1}-T^{1 \underline n}}{\pi^2 {\underline n}^2}\right)^2\\ 
    &= \frac{-20 \sqrt{3} \left(-14 \zeta (3)+36-32 \log (2)+\pi ^2 (\log (256)-5)\right)+45-30 \pi
   ^2+6 \pi ^4}{40 \pi ^4} \\
   &\approx 0.0644771 \,.
\end{aligned}\end{equation}
\noindent For comparison $I_0^{11}= \tfrac{\sqrt{3}}{2}$, $I_{00}^{11}= 1-\tfrac{3}{2 \pi ^2}$, and $I = I_{00}^{11}-(I_{0}^{11})^2 = \tfrac{1}{4}-\tfrac{3}{2 \pi ^2}$.  Finally, transforming back into the canonically transforming variable, we find that the effective Lagrangian density is
\begin{equation}\begin{split}
\mathcal{L}_{\text{eff}}=&-\frac{1}{4} F_{\mu\nu}F^{\mu \nu} -\overline{\big(D_\mu \varphi\big)} D^\mu \varphi - \pi^2 \overline \varphi \varphi +\frac{3}{10}\partial_\mu h\partial^\mu h +\partial^\mu h\big(\partial_\mu g-a_\mu\big) \\
&- e^2_\text{eff}\widetilde{I}\left(a_\mu-\partial_\mu g\right)\left(a^\mu-\partial^\mu g\right)\overline \varphi \varphi - e_\text{eff}^2 \widetilde{X} h^2 \overline \varphi \varphi  \,,
\end{split}\end{equation}
\noindent where $e_\text{eff}=eI_{0}^{11}$ is the effective electric charge, $D_\mu = \partial_\mu - ie_\text{eff}a_\mu$ is the canonical covariant derivative, $\widetilde{I} = I/(I^{11}_0)^2$, and $\widetilde X = X/(I^{11}_0)^2$. \\
\indent The effective theory is Maxwell, with a standard gauge-fixing term, coupled in the usual way to an electrically charged scalar $\varphi$ with charge $e_{\text{eff}}=e I_0^{11}$, out to order $e_{\text{eff}}^1$ in the action. If one only considers this leading behaviour in the effective charge of the theory, its dynamics is physically indistinguishable from that of the usual dimensional reduction\footnote{That is, a von Neuman / von Neuman or periodic $\cS^1$ reduction.} case. At $e_\text{eff}^2$ order, however, we find covert symmetry breaking identical to the symmetry breaking originating in coupling to the zero-level Stueckelberg field arising in the term $\left(a_\mu-\partial_\mu g\right)\left(a^\mu-\partial^\mu g\right)\overline \varphi \varphi$.
\\
\indent In a usual dimensional reduction, the zero-level lower dimensional theory inherits the corresponding projection of the higher dimensional symmetries linearly, and this is sufficient to fix the form of the lower dimensional theory. This is not so in the present case because of the non-constant transverse wavefunction zero-mode, and its associated Stueckelberg field. We can write new structures that are invariant under the higher dimensional symmetry using this nonlinearly transforming Stueckelberg field, which are however physically distinct from the structure of the linearly realised theory in the lower dimension. Accordingly, the higher dimensional symmetry becomes nonlinearly realised in the lower dimension. By explicitly calculating the effective theory, however, we find linear symmetry breaking only appears in a `covert' way, starting at  $a^2|\varphi|^2$ in the action or $|\varphi|^6$ order in scalar only physical processes.
\section{Conclusion and Outlook}

In this paper, we have focused on what we considered to be the simplest case in which covert symmetry breaking reveals itself. This was stimulated by observation of the explicit structure \cite{Crampton:2014hia} of an effective lower-dimensional theory of gravity with a noncompact transverse space, but localised in the lower dimension thanks to a mass gap in the spectrum of the associated Schr\"odinger problem. Clearly, a return to that system needs to be made to carry out a similar investigation to that of this paper. Along the way, an analogous study of pure Yang-Mills theory in $d+1$ dimensions with the Dirichlet/Robin boundary conditions considered here can be done \cite{EHLS}. 

More generally, one also needs to consider what is the best way to approach the evaluation of an effective gravitational theory in a lower dimension when the transverse space is noncompact. The key problem in such cases is the vanishing of the effective Newton constant, 
as pointed out originally in Ref.\ \cite{Hull:1988jw}. There is, however, one known way to get nontrivial interactions in a number of such cases: restrict attention to pure gravity in the lower dimension, or, in the case of a supersymmetric theory, restrict attention to pure supergravity with unbroken supersymmetry. For example, there are lower-dimensional supersymmetric braneworld constructions where such pure supergravity on the brane worldvolume exists as a consistent reduction from the higher dimensional theory \cite{Brecher:1999xf,Chamblin:1999by,Lu:2000xc}. For such pure lower-dimensional supergravity solutions, there really is no clearly defined Newton constant -- for example, any Ricci-flat metric in the lower dimension will continue to give a solution to the higher dimensional field equations. A related feature of such lower-dimensional systems is that they retain a `trombone' symmetry of the lower dimensional field equations, as do all pure supergravity theories. A clear meaning to a gravitational coupling constant arises only when one couples to fields outside the lower-dimensional supergravity supermultiplet. An example of such coupling could be to another kind of braneworld supermultiplet -- branewaves arising as Goldstone modes from broken symmetries of a background brane solution. In such cases, with an infinite transverse space, the problem of a vanishing Newton constant is likely to recur: the branewave modes may couple directly only to the non-zero-level modes of the higher dimensional theory. 

The kind of system investigated in this paper and in Ref. \cite{Crampton:2014hia} with a zero-level transverse wavefunction which has nontrivial dependence on the transverse dimensions can guarantee a nonvanishing interaction coupling constant. One then also needs to consider what the physical implications of the resulting covert style of symmetry breaking might be.

\section*{Acknowledgments}

We are grateful to Stanley Deser, Carl Bender and Jean-Luc Lehners for helpful discussions. KSS would like to thank Cal Tech and the Albert Einstein Institute for hospitality at times during the course of the work. The work of KSS was supported in part by the STFC under Consolidated Grant ST/P0000762/1, the work of ADH was supported by an STFC PhD studentship, and the work of CWE was supported by the United States Department of Veterans Affairs under the Post 9/11 GI Bill.

\section*{Appendices}

\appendix

\section{Maxwellian Degrees of Freedom and Hamiltonian}
\noindent Within this appendix our aim is to provide a detailed account of the physical degrees of freedom and the Hamiltonian for the massless sector of the system that arises in Section \ref{MaxwellSection}. To do this we begin by using the gauge symmetry of the $\varphi_{i}$, (\ref{eq:u1stueckelberg}), to fix $\varphi_{2}$ to zero. Within this gauge, the equations of motion arising from (\ref{eq:massless4dsystem2stueckelbergs}) are
\begin{equation}
\label{eomcoordiateform}
    \Box_{d}A_{\mu}+k\partial_{\mu}\varphi = 0 \hspace{2mm},\hspace{2mm} \partial^{\mu}A_{\mu} = 0 \hspace{2mm},\hspace{2mm} \Box_{d}\varphi = 0\,,
\end{equation}
where we have relabelled $a_{\mu}$ as $A_{\mu}$, and $\varphi_{1}$ as $\varphi$. \\
\indent If we take a Fourier transform of (\ref{eomcoordiateform}), and perform the decomposition
\begin{equation}
    \label{Atildedecomp}
    \tilde{A}_{\mu}(p)= \tilde{\lambda}(p)p_{\mu}+\tilde{a}_{\mu}(p)\,,
\end{equation}
where $\tilde{A}_{\mu}$ is the Fourier transform of $A_{\mu}$ and $p_{\mu}$ and $\tilde{a}_{\mu}$ are assumed to be linearly independent vectors at the momentum-space point $p_\mu$, then we obtain the equations
\begin{equation}
\label{Atildemueom}
-p^{2}\tilde{\lambda}p_{\mu}-p^{2}\tilde{a}_{\mu}-ikp_{\mu}\tilde{\varphi} = 0\,,
\end{equation}
\begin{equation}
    \label{Atildemugauge}
    \tilde{\lambda}p^{2}+\tilde{a}_{\mu}p^{\mu} =0\,,
\end{equation}
\begin{equation}
    \label{phitildeeom}
    p^{2}\tilde{\varphi} = 0\,,
\end{equation}
where $\tilde{\varphi}$ denotes the Fourier transform of $\varphi$. Note if we shift $\tilde{\lambda}$ to $\tilde{\lambda} + \hat{\lambda}$, in (\ref{Atildedecomp}), then (\ref{Atildemueom})-(\ref{phitildeeom}) are invariant if $supp(\hat{\lambda})=\{p_{\mu}|p^{2}=0\}$. \\
\indent We begin by noting that the linear independence of $p_{\mu}$ and $\tilde{a}_{\mu}$ means that (\ref{Atildemueom}) implies
\begin{equation}
    \label{atildeeom}
    p^{2}\tilde{a}_{\mu}=0\,,
\end{equation}
\begin{equation}
    \label{pmupartofAmueom}
    -p^{2}\tilde{\lambda}-ik\tilde{\varphi}=0\,.
\end{equation}
Using (\ref{phitildeeom}) and (\ref{atildeeom}) it follows that 
\begin{equation}
    supp(\tilde{a}_{\mu})=supp(\tilde{\varphi})=\{p_{\mu}|p^{2}=0\}\,,
\end{equation}
which, along with (\ref{Atildemueom}), evaluated when $p^{2}=0$, but where $p_{\mu}\neq0$, gives 
\begin{equation}
    \label{varphisupport}
    supp(\tilde{\varphi})=\{p_{\mu}=0\}\,,
\end{equation}
hence showing this field doesn't correspond to a propagating degree of freedom. This can then be used in (\ref{pmupartofAmueom}) to show that
\begin{equation}
    \label{lambdasupport}
    supp(\tilde{\lambda})=\{p_{\mu}|p^{2}=0\}\,,
\end{equation}
meaning $\tilde{\lambda}$ only has support on the lightcone.\footnote{Which is precisely where we can freely shift this function while keeping the equations (\ref{Atildemueom})-(\ref{phitildeeom}) invariant.}\\
\indent Owing to (\ref{Atildedecomp}) and the fact that $\tilde{\lambda}$ only has support on the lightcone, we find that 
\begin{equation}
    \label{Amupmu}
    p^{\mu}\tilde{A}_{\mu}=p^{\mu}\tilde{a}_{\mu} =0\,,
\end{equation}
where the first equality follows from (\ref{Atildemugauge}) by using (\ref{lambdasupport}). Since we can shift $\tilde{\lambda}$, precisely on its support set, and leave (\ref{Atildemueom})-(\ref{phitildeeom}) invariant, we can set 
\begin{equation}
    \label{lambdainvarinacefixingchoice}
    \tilde{\lambda}=-\frac{\tilde{a}_{0}}{p_{0}}\,,
\end{equation}
on the lightcone, except at $p_{\mu}=0$. This has the effect of setting $\tilde{A}_{0}=0$ on the lightcone, except at $p_{\mu}=0$. This results in (\ref{Amupmu}) leading to the condition
\begin{equation}
    \label{Aipi}
    \tilde{A}_{i}p_{i}=0\,,
\end{equation}
which confirms that the system described by (\ref{eq:massless4dsystem2stueckelbergs}) possesses only $d-2$ propagating degrees of freedom. As a result of this analysis, we see that the system is equivalent to standard Maxwell theory, once we go on shell.

Another way to look at the dynamics of the zero-level system \eqref{eq:lorenzgaugemaxwell} including the Lagrange multiplier field $\Psi_1$ is to consider its Hamiltonian formulation. The inclusion of this field, which pre-selects the Lorenz gauge for $a_\mu$, leads to a modified Hamiltonian formulation since there is no longer an unrestricted $\lambda(x)$ gauge symmetry. This gives rise to a conjugate momentum to $a_0$, \ie $\pi_0=k\Psi_1$, which is not ordinarily present. The canonical action becomes
\be
I_{\hbox{canon}}=\int dt \int d^{d-1}x\left( \pi_i\dot a_i+\pi_0 \dot a_0 - ({\cal H}_t + {\cal H}_v)\right)\,, \quad i=1,\ldots,d-1\label{eq:canact}
\ee
where 
\bea
{\cal H}_t&=&\frac12\pi_i\pi_i+\frac14 F_{ij}F_{ij}\label{eq:Ht}\\
{\cal H}_v&=&\pi_i\partial_ia_0+\pi_0\partial_ia_i\label{eq:Hv}\ .
\eea
Here, ${\cal H}_t$ is the usual positive semidefinite Maxwell Hamiltonian density while ${\cal H}_v$ is a separate quantity whose spatial integral $Q_v=\int d^{d-1}x\, {\cal H}_v$ is independently conserved in time by virtue of the field equations for the canonical action \eqref{eq:canact}. As usual, Noether's theorem relates such a conserved quantity to a global symmetry and here that symmetry is:
\bea
\delta a_i&=& \partial_ia_0\rho\qquad\qquad\ \  \delta \pi_i = \partial_i\pi_0\nn\\
\delta a_0&=& (\partial_ia_i-\pi_0)\rho\qquad  \delta \pi_o = \partial_i\pi_i\rho\,,\label{eq:newmaxsym}
\eea
where $\rho$ is a spacetime-constant parameter. The conserved quantity $Q_v$ is of indefinite sign, but this does not imply the presence of ghost degrees of freedom; the conserved energy can be considered to be just $E=\int d^{d-1}x\, {\cal H}_t$, which is positive semidefinite.
It is helpful to consider what happens to $Q_v$ in a standard Maxwell theory presentation without $\pi_0$: one finds $Q_v=0$ using the usual Gauss's law $\partial_i\pi_i=\partial_i F_{0i}=0$ for noninteracting Maxwell theory. The symmetry \eqref{eq:newmaxsym} is still there (setting $\pi_0\to0$), but it is then a symmetry with a vanishing charge, somewhat reminiscent of the vanishing-charge symmetries of supersymmetric theories without auxiliary fields.

\section{Details of the Commuting Square Diagram for Scalar QED}\label{sec:comsqdiagdtls}
In this appendix, we give details of the equivalence between higher and lower dimensional presentations of the scalar QED dynamics as represented in Figure \ref{fig:reducsquare} and needed in Subsections \ref{ssec:inthdeqnsbc} and \ref{ssec:intldeqnsbc}.

Starting with \eqref{eq:qedwveqn}, recall that the terms $\Phi\partial_\mu\overline\Phi$ and $\overline\Phi\Phi A_\mu$ obey Dirichlet/Robin boundary conditions, and so can be written as linear combinations of the $\{\xi_i(z)\}$ basis given in \eqref{eq:expansionthetasxis}:
\begin{equation}
\theta_n(z)\theta_m(z) = I_i^{nm}\xi_i(z) \,,\quad \theta_n(z)\theta_m(z)\xi_j(z)=I_{ij}^{nm}\xi_i(z) \,.
\end{equation} 
With these overlap integrals, we can use linear independence to read off the lower-dimensional equations coming from \eqref{eq:qedwveqn}. We have 
\begin{equation}
\Box_d a_\mu^{(0)} - \partial_\mu\partial^\nu a_\nu^{(0)}-\partial_\mu h + ieI^{nm}_0\big(\phi^{(n)}\partial_\mu\overline\phi^{(m)}-\overline\phi^{(n)}\partial_\mu\phi^{(m)}\big) - 2e^2I_{0i}^{nm}\overline\phi^{(n)}\phi^{(m)}a_\mu^{(i)} = 0 \,, \label{eq:qedloweramu0}
\end{equation} 
and 
\begin{equation}
\big(\Box_d -\omega_i^2\big)a_\mu^{(i)} - \partial_\mu\partial^\nu a_\nu^{(i)}+\omega_i^2\partial_\mu g^{(i)} + ieI_i^{nm}\big(\phi^{(n)}\partial_\mu\overline\phi^{(m)}-\overline\phi^{(n)}\partial_\mu\phi^{(m)}\big) - 2e^2I_{ij}^{nm}\overline\phi^{(n)}\phi^{(m)}a_\mu^{(j)} = 0 \,, \label{eq:qedloweramui}
\end{equation} 
for $i\in\{1,2,\dots\}$. \\
\indent For \eqref{eq:qedscalareqn}, we define the overlap integrals 
\begin{align}
&P^{nm}=\int_0^1dz\,\zeta^2(z)\theta_n(z)\theta_m(z) \,,\quad Q^{nm}_{i}=\int_0^1dz\,\zeta(z)\xi_i'(z)\theta_n(z)\theta_m(z) \,,\nonumber \\
&T^{nm} = \int_0^1dz\,\zeta(z)\theta_m'(z)\theta_n(z) \,,\quad U^{nm}_i=\int_0^1dz\,\xi_i'(z)\theta_m'(z)\theta_n(z) \,,\nonumber\\
&R^{nm}_{ij}=\int_0^1dz\,\xi_i'(z)\xi_j'(z)\theta_n(z)\theta_m(z) \,.
\end{align} 
Using these, the lower-dimensional complex scalar equations are 
\begin{align}
&\big(\Box_d\phi^{(n)}-m_n^2\phi^{(n)}\big) - ie\big(I_i^{nm}\partial^\mu a_\mu^{(i)} + I_0^{nm}h - I_i^{nm}\omega_i^2g^{(i)}\big)\phi^{(m)} -2ieI_i^{nm}a_\mu^{(i)}\partial^\mu \phi^{(m)} \nonumber \\
&-2ie\big(T^{nm}h+U^{nm}_ig^{(i)}\big)\phi^{(m)}-e^2\big(I_{ij}^{nm}a_\mu^{(i)}a^{(j)\mu} + P^{nm}h^2+ 2Q^{nm}_i h g^{(i)} + R_{ij}^{nm}g^{(i)}g^{(j)}\big)\phi^{(m)} = 0 \,. \label{eq:qedlowerscalar}
\end{align} 
for $n\in\{1,2,\dots\}$. \\
\indent For \eqref{eq:qedzeqn}, it is much more convenient to rewrite the expansion of $A_z$ in terms of the orthonormal basis $\{\psi_\alpha(z)\}$. Defining the overlap integrals 
\begin{equation}
M_\alpha^{nm} = \int_0^1dz\,\psi_\alpha(z)\theta_m'(z)\theta_n(z)\,,\quad N_{\alpha\beta}^{nm}=\int_0^1dz\,\psi_\alpha(z)\psi_\beta(z)\theta_m(z)\theta_n(z) \,,
\end{equation} 
we find the lower-dimensional equations are 
\begin{equation} 
\Box_d\chi_\alpha-c_{i;\alpha}\partial^\mu a_\mu^{(i)} +ieM_\alpha^{nm}\big(\phi^{(n)}\overline\phi^{(m)}-\overline\phi^{(n)}\phi^{(m)}\big) -2e^2N^{nm}_{\alpha\beta}\overline\phi^{(n)}\phi^{(m)}\chi_\beta = 0 \,. \label{eq:qedloweraz}
\end{equation} 
To convert this equation into equations for $h$ and $g^{(i)}$, we contract it with operators $D_{\alpha\beta}b_\beta$ and $D_{\alpha\beta}c_{i;\beta}$, using the relations \eqref{eq:contractionidentities}. After some manipulation, we arrive at the following equations of motion: 
\begin{align}
&\Box_d h = ie U_{0}^{nm}\big(\phi^{(n)}\overline\phi^{(m)}-\overline\phi^{(n)}\phi^{(m)}\big)-2e^2\big(Q_{0}^{nm}h+R_{0i}^{nm}g^{(i)}\big)\phi^{(n)}\overline\phi^{(m)} \,, \label{eq:qedheqn} \\
&\Box_d g^{(0)}-\partial^\mu a_\mu^{(0)} = ie\widetilde T^{nm}\big(\phi^{(n)}\overline\phi^{(m)}-\overline\phi^{(n)}\phi^{(m)}\big)-2e^2\big(\widetilde{P}^{nm}h +\widetilde{Q}^{nm}_ig^{(i)}\big)\phi^{(n)}\overline\phi^{(m)} \,, \label{eq:qedg0eqn} \\
&\omega_i^2\big(\Box_d g^{(i)}-\partial^\mu a_\mu^{(i)}\big) = -ieU_{i}^{nm}\big(\phi^{(n)}\overline\phi^{(m)}-\overline\phi^{(n)}\phi^{(m)}\big)+2e^2\big(Q_{i}^{nm}h+R_{ij}^{nm}g^{(j)}\big)\phi^{(n)}\overline\phi^{(m)} \,, \label{eq:qedgieqn} 
\end{align} 
where $i\in\{1,2,\dots\}$ in \eqref{eq:qedgieqn}, and $\widetilde T^{nm}=T^{nm}-\tfrac{3}{5}U^{nm}_0$, $\widetilde P^{nm}=P^{nm}+\tfrac{3}{5}Q_0^{nm}$, and $\widetilde Q^{nm}_i=Q^{nm}_i+\tfrac{3}{5}R^{nm}_{0i}$. Equations \eqref{eq:qedloweramu0}-\eqref{eq:qedlowerscalar} and \eqref{eq:qedheqn}-\eqref{eq:qedgieqn} are the lower-dimensional equations of motion.\footnote{It is important to note that these equations are internally consistent, as all Bianchi identities are satisfied.} \\
\indent It is a straightforward task to check that \eqref{eq:qedlowerac} produces the same lower-dimensional equations of motion. 
\section{Effective Theory in the Original Variables}
\indent At the end of Section \eqref{sc:ifo} we stated that the system in the original (gauge covariant) higher dimensional variables retains gauge covariance (or invariance at the level of the action) after integrating out all of the (more) massive matter scalars. We described in broad strokes the details of how this occurs, specifically that the action is augmented by new terms at quartic order and the transformation is augmented at quadratic order and together these define an unusual but gauge invariant action (or oddly covariant equations of motion). Here we will show how that invariance works at the level of the action for one term, specifically the $a^2\phi^2$ `unusual coefficient' term.\footnote{Here again $a$ is the massless vector and $\phi$ is the lightest matter scalar.}\\
\indent To show the invariance of just this term it is sufficient to only consider only the leading (in fields and derivatives) corrections arising from integrating out the level $\ell>0$ massive matter scalar fields to both the gauge transformation and action. The relevant approximate solutions to the level $\ell>0$ massive matter scalar ($\underline{n} = 2,3,\ldots$) equations of motion are
\begin{equation}
    \phi^{(\underline n)} = \frac{i e}{\pi^2 N^2}\left(\left( 2 a_\mu \partial^\mu \phi + \partial^\mu a_\mu \phi\right) I_{0}^{1\underline n}+ h \phi \left(T^{\underline n 1}-T^{1 \underline n}\right) + 2 g \phi U_0^{\underline n 1}\right) +\cO\left(\Phi^3,{\partial_\mu}^2\right)\;.
\end{equation}
\noindent Here $\Phi$ indicates all corrections arising from recursively putting fields on-shell in their own equations of motion and $\partial_\mu$ indicates arbitrary corrections with more world-volume derivatives, and all integrals ($I$, $T$, and $U$) are as given in  Appendix \ref{sec:comsqdiagdtls}. The new terms in the Lagrangian arising from putting these fields on-shell are
\begin{equation}
    \frac{e^2}{\pi^2 \underline{n}^2}\left| \left( 2 a_\mu \partial^\mu \phi + \partial^\mu a_\mu \phi\right) I_{0}^{1\underline n}+ h \phi \left(T^{\underline n 1}-T^{1 \underline n}\right)+ 2g  \phi U_0^{\underline n 1}\right|^2 + \cO\left(\Phi^5,{\partial_\mu}^2\right) \;.
\end{equation}
\noindent Only two of these new terms are relevant to the terms in the gauge transformation of the action containing one $a$ and two $\phi$:
\begin{equation}
    e^2 2g \overline \phi \left(2 a_\mu \partial^\mu \phi + \partial^\mu a_\mu \phi\right) \frac{I_{0}^{1\underline n}U_0^{\underline n 1}}{\pi^2 \underline{n}^2} + c.c.\;.
\end{equation}
\noindent The relevant terms arising from gauge transforming the above are the terms coming from the transformation of the Stueckelberg field alone:
\begin{equation}
    2 e^2 \lambda \overline \phi \left(2 a_\mu \partial^\mu \phi + \partial^\mu a_\mu \phi\right) \frac{I_{0}^{1\underline n}U_0^{\underline n 1}}{\pi^2 \underline{n}^2} + c.c.\;.
\end{equation}
\indent Similarly, we recall from \eqref{eq:scalaru1} that the lightest scalar field transforms under gauge transformations into scalar fields at all levels, so when we put the heavy fields on-shell we must also put them on-shell in the lightest field's gauge transformation, 
\begin{equation}
    \delta \phi = i e \lambda \phi I_{0}^{11} + e^2 \lambda \left( 2 a_\mu \partial^\mu \phi + \partial^\mu a_\mu \phi\right) \frac{I_{0}^{1 \underline n } I_{0}^{\underline n 1}}{\pi^2 \underline{n}^2} + \cO\left(h,g,\Phi^3,{\partial_\mu}^2\right)\;.
\end{equation}
\noindent The above term quadratic in fields will generate, when substituted into the $\phi$'s mass term, terms with one gauge parameter, one gauge field, and two matter scalars. Specifically the correction is
\begin{equation}
    \delta \left( - \pi^2 \left|\phi\right|^2\right) = \ldots - \pi^2 \overline \phi \left(e^2 \lambda \left( 2 a_\mu \partial^\mu \phi + \partial^\mu a_\mu \phi\right) \frac{I_{0}^{1 \underline n} I_{0}^{\underline n 1}}{\pi^2 \underline{n}^2}\right) + c.c. + \ldots \;.
\end{equation}
\indent Lastly, we remember that the coefficient of the quartic term is ``unusual'' because it is not the anticipated square of the cubic term's coefficient. Taking the transformations of these two terms together, we collect only the term which which contains one gauge parameter, one gauge field, and two matter scalars:
\begin{equation}
    \delta\left(-ie a_\mu \left(\overline \phi \partial^\mu \phi - \phi \partial^\mu \overline \phi\right) I_{0}^{11} - e^2 a_\mu a^\mu \overline \phi \phi I_{00}^{11}\right) = \ldots-2 e^2 a_\mu \partial^\mu \lambda \overline\phi \phi\left(I_{00}^{11}-{I_{0}^{11}}^2\right)+\ldots\;.
\end{equation}
These are all the terms in the gauge variation of the Lagrangian that are of the `$\partial \lambda a \overline \phi \phi$' variety. If we take all the terms that we've detailed above and integrate by parts we find that they may be written as
\begin{equation}
     -2 e^2 a_\mu \partial^\mu \lambda \overline\phi \phi\left(\left(I_{00}^{11}-{I_{0}^{11}}^2\right)-\pi^2 \frac{I_{0}^{1\underline n }I_{0}^{\underline n 1}}{\pi^2 \underline{n}^2}+ 2\frac{I_{0}^{1\underline n }U_0^{\underline n 1}}{\pi^2 \underline{n}^2}\right) \;.
\end{equation}
\noindent For the Lagrangian to be gauge invariant the coefficient of the above term must vanish, or
\begin{equation}\begin{gathered}
    I=\int_0^1 \xi^2 \theta^2 dz - \left(\int_0^1 \xi \theta^2 dz\right)^2 - \sum_{N=2}^\infty \frac{\pi^2}{\pi^2 N^2}\int_0^1 \xi(s) \theta(s) \theta_N(s) ds \int_0^1\xi(z) \theta(z) \theta_N (z) dz\\
    +\sum_{N=2}^\infty\frac{2}{\pi^2 N^2} \int_0^1 \xi'(s)\theta'(s) \theta_N(s) ds \int_0^1 \xi(z) \theta(z) \theta_N(z) dz =0\;.
\end{gathered}\end{equation}
\indent In order for the above Fourier basis, each of these integrals is known.\footnote{Each is done by repeated integration by parts.} The resulting sums are also doable
\begin{equation}
    I = 1 - \frac{3}{2 \pi^2} - \left(\frac{\sqrt{3}}{2}\right)^2 - \frac{48}{\pi^4} \sum_{N=2}^\infty \frac{(1+(-1)^2)^2}{(n^2-1)^4}+\frac{48}{\pi^5}\sum_{N=2}^\infty\frac{(1+(-1)^2)^2}{(n^2-1)^3}=0\;.
\end{equation}
\indent To summarise, we have, for the effective theory in the original variables, gauge transformed, then collected all terms including one power of the gauge parameter, one power of the gauge field, two powers of the scalar, and one world-volume derivative and have shown that these terms sum to zero. While this only shows the invariance in the action of a single term, it is torturous enough to calculate this. Furthermore, we know that these variables are simply a field redefinition away from the more easily manifestly gauge invariant variables used in Section \eqref{sc:tfet}, so the final action expressed in either set of variables proves to be invariant.

\end{document}